# Fake news propagate differently from real news even at early stages of spreading

**Zilong Zhao[1,2], Jichang Zhao[3], Yukie Sano[4], Orr levy[5], Hideki Takayasu[6,7], Misako Takayasu[7], Daqing Li[1,2,*], Junjie Wu[3,8,*], Shlomo Havlin[5,7]**

[1]School of Reliability and Systems Engineering, Beihang University, Beijing 100191, China

[2]Science and Technology on Reliability and Environmental Engineering Laboratory, Beijing 100191, China

[3]School of Economics and Management, Beihang University, Beijing 100191, China

[4]Faculty of Engineering, Information and Systems, University of Tsukuba, Tennodai, Tsukuba, Ibaraki 305-8573, Japan

[5]Department of Physics, Bar-Ilan University, Ramat Gan 5290002, Israel

[6]Sony Computer Science Laboratories, 3-14-13 Higashi-Gotanda, Tokyo 141-0022, Japan

[7]Institute of Innovative Research, Tokyo Institute of Technology, Yokohama 226-8502, Japan

[8]Beijing Advanced Innovation Center for Big Data and Brain Computing, Beihang University, 100191, Beijing, China.

Zilong Zhao, Jichang Zhao, Yukie Sano，and Orr Levy contributed equally to this paper.

* To whom correspondence may be addressed daqingl@buaa.edu.cn and

wujj@buaa.edu.cn .

# Abstract

Social media can be a double-edged sword for society, either as a convenient channel exchanging ideas or as an unexpected conduit circulating fake news through a large population. While existing studies of fake news focus on theoretical modeling of propagation or identification methods based on machine learning, it is important to understand the realistic mechanisms between theoretical models and black-box methods. Here we track large databases of fake news and real news in both, Weibo in China and Twitter in Japan from different culture, which include their complete traces of re-postings. We find in both online social networks that fake news spreads distinctively from real news even at early stages of propagation, e.g. five hours after the first re-postings. Our finding demonstrates collective structural signals that help to understand the different propagation evolution of fake news and real news. Different from earlier studies, identifying the topological properties of the information propagation at early stages may offer novel features for early detection of fake news in social media.

# Introduction

Social networks such as Twitter or Weibo, involving billions of users around the world, have tremendously accelerated the exchange of information and thereafter have led to fast polarization of public opinion [1]. For example, there is a large amount of fake news about the 3.11 earthquake in Japan, where about 80 thousand people have been involved in both diffusion and correction [2]. These fake news, which can be fabricated stories or statements yet without confirmation, circulate online pervasively through the conduit offered by on-line social networks. Without proper debunking and verification, the fast circulation of fake news can largely reshape public opinion and undermine modern society [3]. Even worse, fake news can be intentionally fabricated, leading to diverse threats to modern society including turmoil or riot. The later fake news is revealed and corrected the greater the damage it can make, due to its fast propagation. Thus, detecting fake news at their early stages, in order to effectively avoid further risks and damages, is crucial.

Different from the age of word of mouth, identification of fake news in online social network by experts is generally labor-intensive with low efficiency and accuracy [4], which has attracted much research attention to provide alternative solutions. One intuitive idea for understanding fake news spreading is inspired by epidemic models. In 1960s, Daley and Kendall proposed the so-called DK model [5] in which agents are divided into ignorant, spreader and stifle. Its later extensions include the known epidemic spreading models such as SIS model [6, 7], SIR model [8, 9], SI model [10, 11] and SIRS model [12]. While these studies focus on theoretical modeling of fake news propagation, availability of real data in online social platforms, as we show here, can provide an opportunity to deepen our understanding for the realistic information cascades. Different kinds of observations have been made in empirical studies of fake news, including linguistic features [13], temporal features of re-postings [14-16] and user profiles [17-19]. Actually, information cascades in online social networks are collective propagation networks of which critical topological features remain yet unknown. This motivates our present study to analyze and compare empirically between propagation networks of fake and real news, especially in their early stage, so as to identify the propagation differences and mechanisms behind. These topological features could help to design machine learning approaches to essentially boost the accuracy of fake news targeting [20-22].

Very recently, based on empirical datasets, it has been found that the propagation network of fake news is different from that of real news [23]. They have found that falsehood propagates significantly farther, faster, deeper, and broader than truth news in many categories of information. While this study provides the possibility to differentiate fake news from real news based on propagation network, it remains unclear how this difference between fake news and real news emerge and how soon one can separate these two types. Thus, a systematic study for dynamic evolution of propagation topology is still missing. This motivated us to explore deeper in this direction of how the propagation evolves topologically in different scenarios. With collected real data, we identified early signals for identifying fake news, at five hours from the first re-posting, without other information on contents or users. Note that different from considering all the cascade components [23], our finding is valid for even only following the largest cascade component.

Based on realistic traces of real and fake news propagation in both Weibo (from China) and Twitter (from Japan), we use the re-posting relationships between different users to establish propagation networks. For definition and sampling methods see Methods. With topological signals at different scales, we find that fake news show significant differences from real news. These novel topological features will enable us to design an efficient algorithm to distinguish between fake news and real news even shortly after their birth.

## Results

To construct propagation network of fake and real news, we utilize the re-posting relation between different users participating in circulating the same message (*see Methods*). A schematic description of such propagation networks is shown in Fig. 1A. Typical propagation networks of fake news and real news in Weibo and Twitter are demonstrated in Fig. 1B-E. The topology of propagation network of fake news and real news can be seen to be different. For example, the number of layers in fake news (*Fig. 1B and 1D*) is typically larger than that of real news (*Fig. 1C and 1E*). Additionally, from looking at various examples of fake news propagation networks, it is somewhat surprising that for widely distributed fake news, the creator does not usually have the largest degree in the propagation network (*Fig. S1 and S2*). In the following, our analysis considers also real news created by non-official sources, to avoid the artificial differences due to different types of information creators (official or non-official accounts).

**Layer ratio.** The layer number is defined as the numbers of hops from the creator for a given propagation network. The cumulative numbers of nodes at different layers as a function of time for four typical networks of fake news (*Fig. 2A for Weibo and 2C for Twitter*) and real news (*Fig. 2B for Weibo and 2D for Twitter*) are demonstrated. The fraction of re-postings in the first layer of fake news network is found significantly smaller than that of real news, while the fraction in other layers for fake news is significantly larger than that of real news. Early adopters re-posting the message shortly after the creator play dominant role in circulating real news comparatively. These different roles lead to distinctive landscapes of propagation networks.

The investigation of layer sizes in propagation networks demonstrated in Fig. 2, are systematically extended to all the available messages. As shown in Fig. 3A and 3B, fake news networks tend to possess a relatively smaller first layer, while other layers are larger comparatively. Therefore, we can define the ratio of layer size as the ratio between size of the second and the first layer. As shown in ratio distribution (Fig. 3C and 3D), the ratio in fake news is significantly larger than that of real news. The distribution for ratio of layer sizes separates fake and real news well with only a small overlapped area. Furthermore, it is seen in Fig. 3C that this difference is already significant only at five hours since the first re-posting. In Fig. 3D, it is seen that, for the whole lifespan, the separation of the fake and the real is also significant. In circulation of fake news, the success of the propagation depends highly on the branching process creating different layers, which show different evolution paths between fake and real news. We further investigate the probability difference between fake and real news based on distributions of layer ratio from the time of first re-posting (*Fig. S3 and S4*). Note that the layer size distribution has a peak around layer four in twitter in Fig. 3B, probably due to secondary outbreaks.

It should be noted that while ordinary users post fake news, real news is more probable to be created by official accounts such as government agencies or mass media. In order to eliminate the possible impact of official creators, we also investigate the distribution of ratio of layer sizes in real news with non-official creators. While official news and non-official news have different sample size here, we found they share similar propagation patterns. For example, in Fig. 3C and 3D, all the real news and non-official real news are found to have close distribution of layer size ratio. To verify our results, we also analyze data of 2000 more real news with non-official accounts from another dataset shown in Fig. S5 and S6. The distributions of real news in different dataset are similar yet distinct from that of fake news.

**Characteristic distance.** While the ratio of layer sizes can be regard as a local feature of the network structure, we further inspect a global feature in terms of characteristic distance in a propagation network. As seen in Fig. 4A, distances between pairs of nodes in fake news are in general longer than those of real news, implying that later adopters foster the penetration of fake news in social networks. In order to quantify this finding for all the networks, we propose a second measure called characteristic distance (a) shown in Fig. 4B (*see Methods*). Considering the characteristic distance of all the networks as in Fig. 4B, fake news possesses a significantly longer characteristic distance (4.26) than that of real news (2.59). Similar results can also be observed in Twitter propagations (*Fig. 4C*). The distributions of characteristic distances for all networks are shown in Fig. 4D, where the two curves of fake and real news are well separated. Different from results in [23], we found the size distributions between fake and real news could be similar (*Fig. S7*), while the characteristic distance is significantly different between fake news and real news. We also verified that propagation size has few correlation with the characteristic distance (*Fig. S8*). To verify our results, we also analyze data of 2000 more real news from another dataset shown in Figs. S5.

**Structural Heterogeneity**. Network topology describes the geometry of connections, with more information embedded than the scale statistics in [23]. Here we measure the Heterogeneity (*see Methods*) between propagation networks in fake and real news. The parameter *h* reflects the difference between a given propagation network and its counterpart of a star network with the same size. Network with smaller *h* means similar to star network. Although the out-degree distribution demonstrates only minor difference between fake news and real news (*Fig. S9*), it is interestingly found here that the topology heterogeneity is significantly distinguishable. Note that the relationship between heterogeneity and *N* for star networks is power-law as seen in Fig. 5A. The *h* is the difference between the logarithm of a real network heterogeneity value $H_r$ and the logarithm of heterogeneity value of the same size star network $H_s$. The parameter *h* of fake news is significantly larger compared to that of real news. Consistent findings can also be observed in Twitter (Fig. 5B). In order to quantify the heterogeneity systematically, two distributions of *h* considering different time intervals are calculated. In Fig. 5C, it shows significant difference at five hours from the first re-posting. For the whole propagation lifespan in Fig. 5D, *h* of fake news is also significantly larger than that of real news. Fake news networks have typically lower heterogeneity (larger *h*) since their propagation involves few dominant broadcasters. On the contrary, real news demonstrates higher heterogeneity (smaller $h$) and more star-like layout. The ability to distinguish fake news from real ones is also valid for real news posted by non-official users (*Fig. S5 and S10*). This implies that the indicator based on degree heterogeneity is independent of the type of creator. Additionally, another measure named Herfindahl-Hirschman Index (HHI [24]), shows also distinction between fake news and real news (*Fig. S11*).

The distinction between fake and real news of the heterogeneity measure is the highest among the above three indicators as seen in Fig. 6 and Table 2. For a given Weibo network, measuring its $h$ provides a clear difference between fake news and real news (Fig. 6A), even only considering re-postings at five hours from the first re-posting. This identification becomes even sharper in Fig. 6B, when we consider all re-postings. We show in Fig. 6C the difference significance (*see Methods*) between fake news and real news for different $h$. The differences are about 76% and 79% respectively for re-postings at a relatively short time (five hours) and all re-postings. Note that the probability of being fake news at five hours is already very similar to that for the whole propagation lifespan. The verification analysis (shown in Fig. S5 and S6) also demonstrates the significant difference between fake news and real news with a different dataset, which is fully published by non-official account. Our results suggest that even without sophisticated features like texts or user profiles, direct and understandable topological features can offer high significance for developing early detections.

## Discussion

Being the most vital and popular form of new media, online social networks, fundamentally enhance the creation and dissemination of fake news [25, 26]. Though existing solutions, especially the inspired machine learning approaches, perform impressively on targeting fake news, their black-box style essentially prevents a solid understanding and corresponding method development of debunking or blocking false information. In the other way, human intensive labor approach is time consuming and expensive. For example, it usually takes at least three days [4] for verification and therefore misses the optimal prevention window before massive spreading. In this sense, novel approaches that could help to identify fake news at early stages are urgently needed in preventing the negative impact of false information on modern society.

We show here that fake news spread with a very different network topology, even at early stages, from authentic messages. We focus, in this manuscript on the evolution differences between the propagation topology of two types of information at early stages rather than providing a comprehensive prediction approach [22]. Even taking only one feature, the difference between fake news and real news is significant. The propagation mechanism, which essentially couples information dynamics and collective cognition in social networks, results in a distinctive landscape of circulations between fake and real news. In this way, several early signals can be derived, including the layer-ratio, the characteristic distance and the heterogeneity. Different from sophisticated and time-consuming features that have been considered in contemporary solutions, our suggested measures are simple, understandable, and time-efficient. For example, the weak heterogeneity of fake news might be the result of opinion competition from weak ties between social communities. As stated that "bad" is usually more influential than "good" [27], the unconsciousness of "negative-bias" might result in late burst of fake news, which essentially differs from the spread of real ones. Disclosing intelligence factors that generate the specific topological features we found here can be a promising research direction in the future.

Note that our study has several major differences from Vosoughi et al. [23]. We focus more on the topological features (shape of network), rather than on scale measures of propagation networks (depth or width). Furthermore, we focus on the largest cascade component of propagation network, while all the cascade components are considered in [23]. As both manuscripts confirm the difference between fake news and real news in different aspects, we surprisingly find that this difference can be very significant even at the early stages of propagation.

# Methods

**Weibo data preprocessing.** We analyze 1,701 fake news of Weibo propagation networks (with 973,391 users) and 492 real news of Weibo propagation networks (with 347,401 users) that spread on Weibo from 2011 to 2016. We choose here large networks with more than 200 tweets. More details are given in Table 1. The topics of these Weibo propagation networks include political fake news, economic fake news, fraudulent fake news, tidbit fake news and pseudoscience fake news. (*Fig. S12*).

The retweeting traces of fake news and real news are thoroughly collected through Weibo's open APIs [28]. Fake news is officially investigated and confirmed by the platform of Weibo [29]. Regarding real news, we collect them directly from reliable Weibo accounts. Creators of the real news can be official accounts, for example, government accounts and on-line newspaper accounts. All these real news accounts have been officially verified by the platform of Weibo. On the other hand, we also select manually 51 out of 492 real news networks whose creators are not official accounts. To verify our results, we analyze more data (2000 more real news) from Weibo in the Fig. S5 and S6. These 2000 real news are collected by the same way as above, while these 2000 real news are not official accounts.

In order to create the network, in which nodes are users of Weibo and links are re-postings, we first mine the following data both for fake and real news:

(a) Users: the unique serial number of users who participate in the same network. We also mark the node of the network creator.

(b) Re-postings: the unique serial number of directed re-posting activities, and the serial number of source users and reposted users of this re-posting.

**Twitter data (a validation dataset) preprocessing.** We analyze 27 fake news of Twitter networks (with 105,335 users) and 28 real news of Twitter networks (with 133,109 users). We choose here large networks with more than 200 tweets. More details are given in Table 1.

Twitter data was collected from Japanese tweets posted during the period between 2011/3/11 and 2011/3/17 during the Great East Japan earthquake period. In this period, much fake news propagated in the Japanese Twitter. For fake news, we first gathered 57 topics listed on the website [30]. The contents include tweets with no evidence and malicious tweets, such as starvation of baby and elderly people, someone under the server rack needed help, and the Japan prime minister was taking luxury supper during the disaster. When collecting tweets, we combine a few keywords related to the contents of each fake news. These keywords were proper nouns, such as place names and personal names. After that, we excluded correction tweets whose contents are against fake news including keywords such as "false" and "mistake". Typical procedure to gather fake news is explained in [2].

For real news, we gathered 71 topics by combining keywords (proper nouns, such as place names and personal names) as the fake news. We collected most of tweets originated from official accounts with verified Twitter badges such as government agencies, major newspapers and famous people. The contents included tweets about earthquake information, traffic information, donation information and so on. In addition, we also collect five topics originated from civilians without badges, which were widely retweeted. These tweet contents were related to small real tips during the disaster.

After gathering fake and real news tweet on a keyword basis, we focused on those with more than 200 tweets to create retweet network. We created retweet networks by using the mention symbol "@" in the tweets as a trigger. Because account name (username) of Twitter cannot be longer than 15 characters and can only contain alphanumeric characters (letters A-Z, numbers 0-9) except for underscores [31], we extracted those strings that consist of 2 to 12 characters after "@" as account names. Note that we skipped the account names of one character because it rarely appeared in Japanese Twitter space, and often appeared as the usage of emoticons such as "@ _ @ (surprised face expression)." Furthermore, since there is no distinction between uppercase and lowercase letters in account name rules, all account names were converted to lowercase letters to proceed.

If there are multiple "@" in one tweet, according to the above rules, we extracted multiple account names as much as possible and linked them in order from the beginning of the sentence to create the networks. Basically, from one tweet, there exists only one account that retweeted previously. However, tweets were often deleted afterwards in particular fake news, and the network was segregated by this approach. Therefore, by going back to clues in the remaining tweets, we create the network by extracting as much accounts as possible from one tweet.

We create networks with the node as an account name and the link as a connection at the mention symbol "@" with the above rules. We extract the largest connected component (LCC) with no consideration of link directions, and analyze only those with LCC size above roughly 200 nodes. Account names with the oldest tweet time in LCC were treated as creators.

**Definition of fake news and real news.** In recent paper by Lazer et al. [32], "fake news" is defined as fabricated information that mimics news media content in form, without news media's editorial norms and processes for ensuring the accuracy and credibility of information. In our manuscript with Weibo data, the fake news is false information fact-checked by the platform and verified as having been fabricated. Regarding real news, we collect them directly from reliable Weibo accounts. And all these real news accounts have been officially verified by the platform of Weibo. For

Twitter data, the fake news is also false information fact-checked by reliable platform [30]. This is similar as the true/false news defined in paper by S. Vosoughi et al. [23]. And their fake news cascades are checked independently by six fact-checking organizations (snopes.com, politifact.com, factcheck.org, truthorfiction.com, hoax-slayer.com, and urbanlegends.about.com).

**Establishing a network model.** Based on the information we analyze above, we establish a directed network as demonstrated in Fig. 1A. The users are the nodes in the network, and the re-postings are the edges in network. And we mark network creator using color green. Each edge has a direction which is either from creator to re-poster or from former re-poster to latest re-poster. After modeling, we could plot figures of typical networks for both fake and real news of Weibo and Twitter as shown in Fig. 1B to 1E.

**Ratio of layer sizes.** The layer number is defined as the numbers of hops of re-postings from the creator. The ratio of layer sizes is a measure for each network defined as:

$$ratio\ of\ layers = \frac{n_2}{n_1} \qquad (1)$$

$n_1$ and $n_2$ are the sizes (number of nodes) of the first and second layer for a certain network respectively.

**Characteristic distances.** In order to measure the distances, for each network we first calculate the distances between all pairs of nodes in the network, and plot the distribution in a logarithmic scale (y axis). It can be seen from Fig. 4 that the function can be approximated by an exponential function. We consider linear part of curves where their x value (distance) is above one. We calculate the characteristic distance ($a$) accordingly:

$$y \sim e^{-\frac{x}{a}+b} \tag{2}$$

**Heterogeneity measure.** The heterogeneity [33] is defined as:

$$\text{Heterogeneity} = \frac{\sqrt{<k^2>}}{<k>} = \frac{\sqrt{\frac{1}{N}\sum_{i=1}^{N}k_i^2}}{\frac{1}{N}\sum_{i=1}^{N}k_i} \tag{3}$$

$N$ : The number of nodes in the network

$k_i$ : The degree of node $i$

We show a scatter plot (Fig. 5A) for both fake and real news of Weibo. The black line is the theoretical line for star network:

$$\text{Heterogeneous} \sim \sqrt{N} \tag{4}$$

The $h$ is the difference between the logarithm of a real network heterogeneity value $H_r$ and the logarithm of heterogeneity value of the same size star network $H_s$ as shown below:

$$h = \log(H_r) - \log(H_s) \tag{5}$$

**Probability of being fake news.** Here we use ratio of layer sizes as an example. We divide ratio of layer sizes into $n$ portion. In the $i$th portion, the probability of being fake news is:

$$p = \frac{p_i^f}{p_i^f + p_i^r} \quad (6)$$

$p_i^f$: The probability of fake news in the $i$th portion (the number of fake news in this portion divided by the total number of fake news).

$p_i^r$: The probability of real news in the $i$th portion.

**Significance of difference.** When we distinguish fake news from real ones using topological measures such as ratio of layer sizes or the characteristic distance, it is important to know the significance of difference. Here we use ratio of layer sizes as an example. First, we rank the Weibo propagation networks by their ratio of layer sizes ignoring their types (fake or real). Second, we randomly split these propagation networks into $n$ portions which have same number of networks. Finally, we calculate the difference significance using following formula:

$$Q = \frac{1}{n} \sum_1^n \frac{\max(p_i^r, p_i^f)}{p_i^r + p_i^f} \quad (7)$$

$n$: The number of portions that we divide.

## Acknowledgements

This work is supported by National Natural Science Foundation of China (Grant No.

71822101). S. H. thanks the Israel Science Foundation, ONR, the Israel Ministry of Science and Technology (MOST) with the Italy Ministry of Foreign Affairs, BSF-NSF, MOST with the Japan Science and Technology Agency, the BIU Center for Research in Applied Cryptography and Cyber Security, and DTRA (Grant no. HDTRA-1-10-1-0014) for financial support. Z. J. was supported by NSFC (No. 71501005) and the National Key Research and Development Program of China (No.2016QY01W0205). This work partly was supported by JSPS KAKENHI Grand Number 17K12783 (Y.S.). H.T. and M.T. are supported by JST Strategic International Collaborative Research Program (SICORP) on the topic of "ICT for a Resilient Society" by Japan and Israel. W.J. was partially supported by National Natural Science Foundation of China (NSFC) (71725002, 71531001, U1636210, 71471009) and Fundamental Research Funds for Central Universities.

## Data availability

Our data are available from the corresponding author on reasonable request.

# Figures

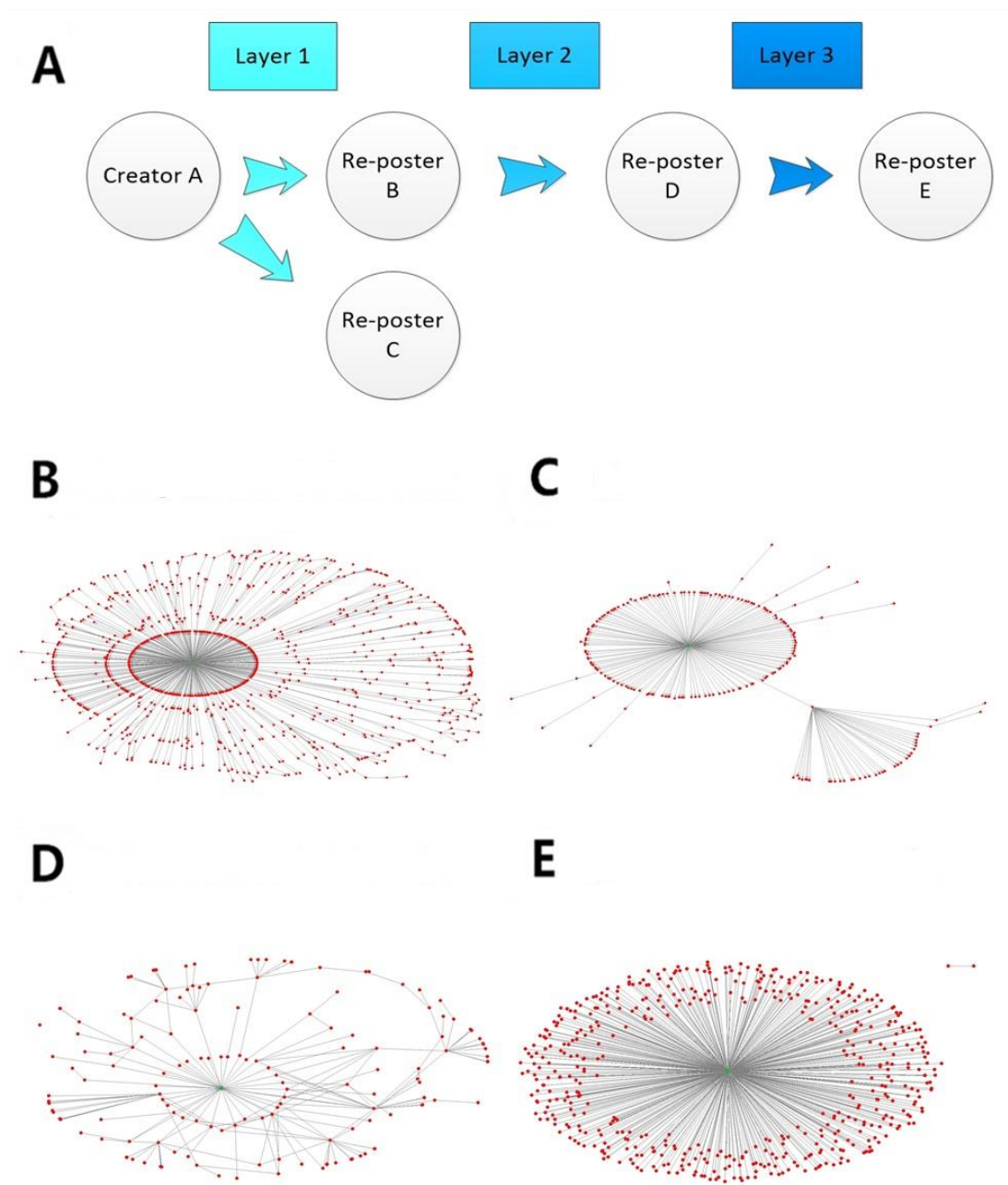

**Figure 1. Typical examples of fake and real news networks.** (A) Schematic diagram of the propagation of a post and its re-posting. The nodes represent the users and the edges are the re-postings. The directionality determines which user is the re-poster among the two users: the origin is the former re-poster and the target is the later re-poster. In this example, there are five users in which one user is the creator, A, who publishes the first post and the other four users repost on this topic. Re-poster B and re-poster C reposts directly from the creator, A, while D reposts C and E reposts D. A layer consists of re-postings whose re-posters have the same distance from the creator. We color the edges according to their layer from light to dark blue. (B) A real typical Weibo network of fake news with 1123 nodes. The edge's arrow stands for its direction. This network of Weibo is a fake news about health problems due to a milk tea shop. (C) A typical Weibo real news network with 215 nodes. This network of Weibo is about a tip of prevent sunstroke. (D) A typical Twitter fake news network with 199 nodes. This tweet is about an electric store that raised the price of a battery unreasonably. (E) A typical real news network in Twitter with 578 nodes. This tweet is a correction tweet against fake news about Cosmo oil by Asahi newspaper. We applied the Fruchterman-Reingold layout by using Pajek software here.

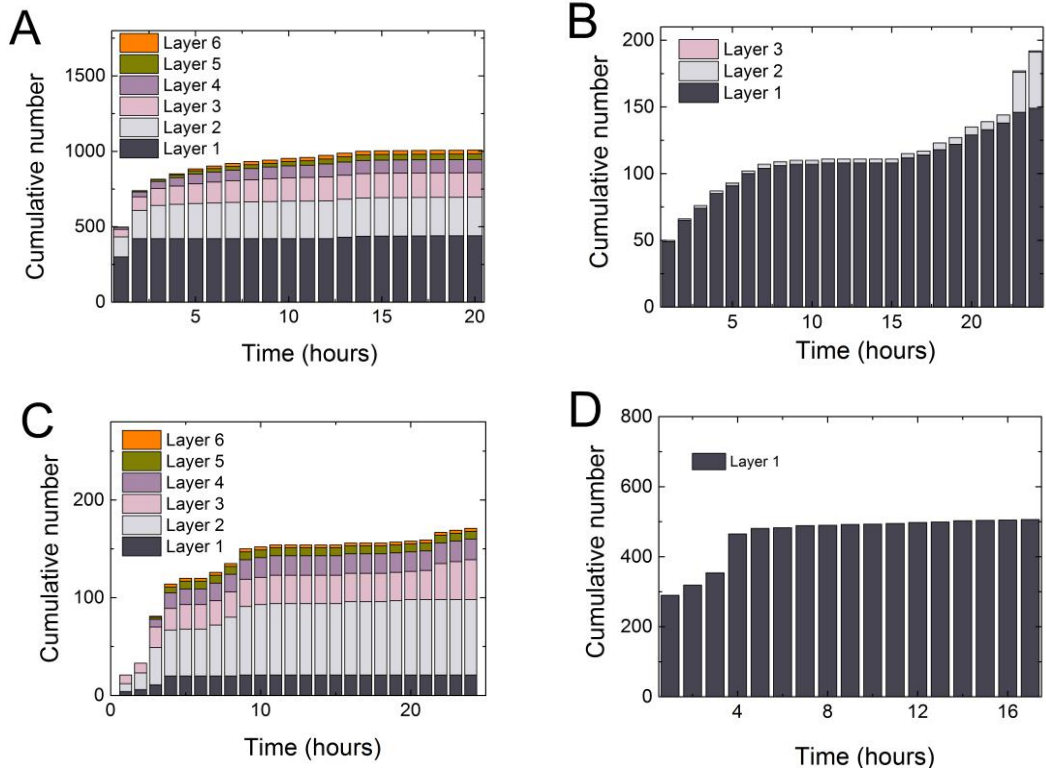

**Figure 2. Different layer sizes grow with time in typical networks.** The y axis is the cumulative number of re-postings at different layers of typical networks in Weibo and Twitter. The x axis is the time (in hours) from the time of news creation and the different colors stand for different layers. Shown examples are (A) fake news and (B) real news in Weibo, as well as (C) fake news and (D) real news in Twitter. In Fig. 2A, the fraction of nodes located in the first layer is around 45% of all the nodes at the end of propagation. However, in Fig. 2B, the total number of nodes in the first layer occupies about 78% of all the nodes at the end of propagation. If the total number of nodes does not change much after 20 hours, we ignore the re-posting after 20 hours in order to clearly see the layers in the figure. These four typical networks are the same networks shown in Fig. 1. It is seen that the layer sizes of real news and fake news are significantly different in both Weibo and Twitter. Real news networks tend to have a relatively larger first layer, while fake news networks are relatively uniformly distributed in different layers.

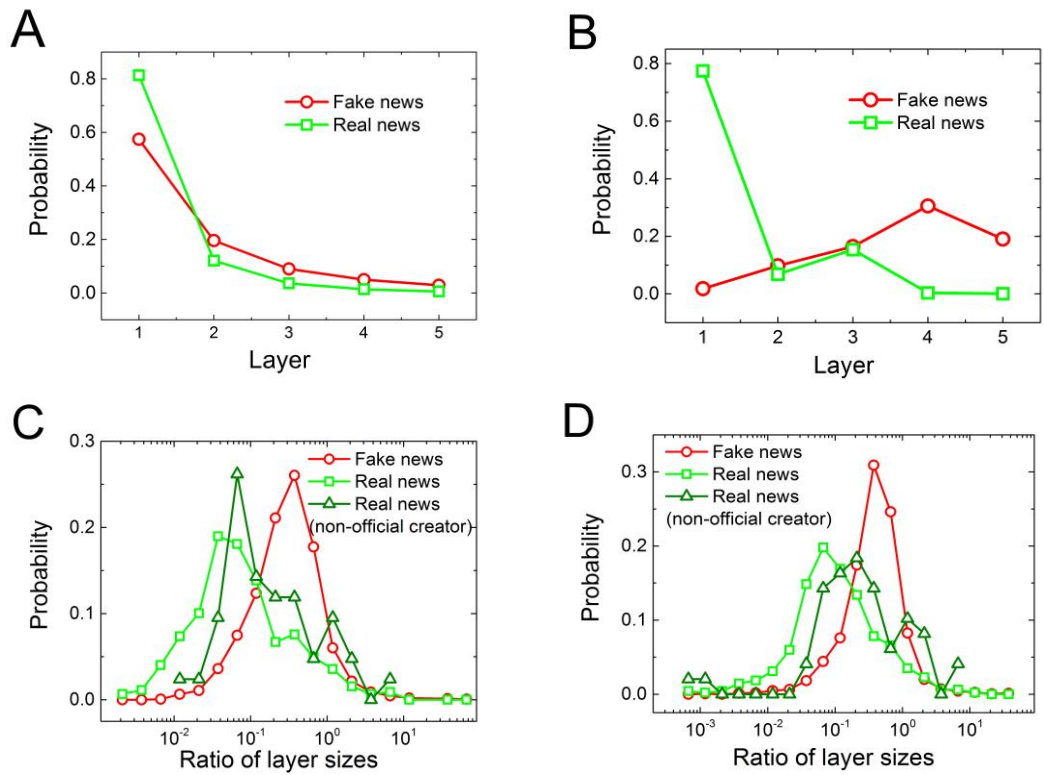

**Figure 3. Ratio of layer sizes differentiates fake news from real news.** The distribution of ratio of layer sizes and its development after a period of time can differentiate fake news from real news. These differences appear already after a few hours. (A) The PDF of all re-postings in the first five layers averaged over all of the Weibo propagation networks. The *p*-value of Mann-Whitney is below 0.01. (B) The PDF of all re-postings in the first five layers in all networks of Twitter. The *p*-value of Mann-Whitney is below 0.01. (C) Distribution of ratio of layer sizes at five hours from the first re-posting. The ratio of layer sizes is the size of the second layer divided by the size of the first layer. The *p*-value of Mann-Whitney between fake and real news is below 0.01. This figure considers 1682 fake news, 492 real news and 51 real news with non-official creator at five hours from the first re-posting. (D) Distribution of ratio of layer sizes of all re-postings for the whole lifespan. The *p*-value between fake and real news is below 0.01. Here we consider all available Weibo propagation networks (1701 fake news, 492 real news and 51 real news with non-official creator networks).

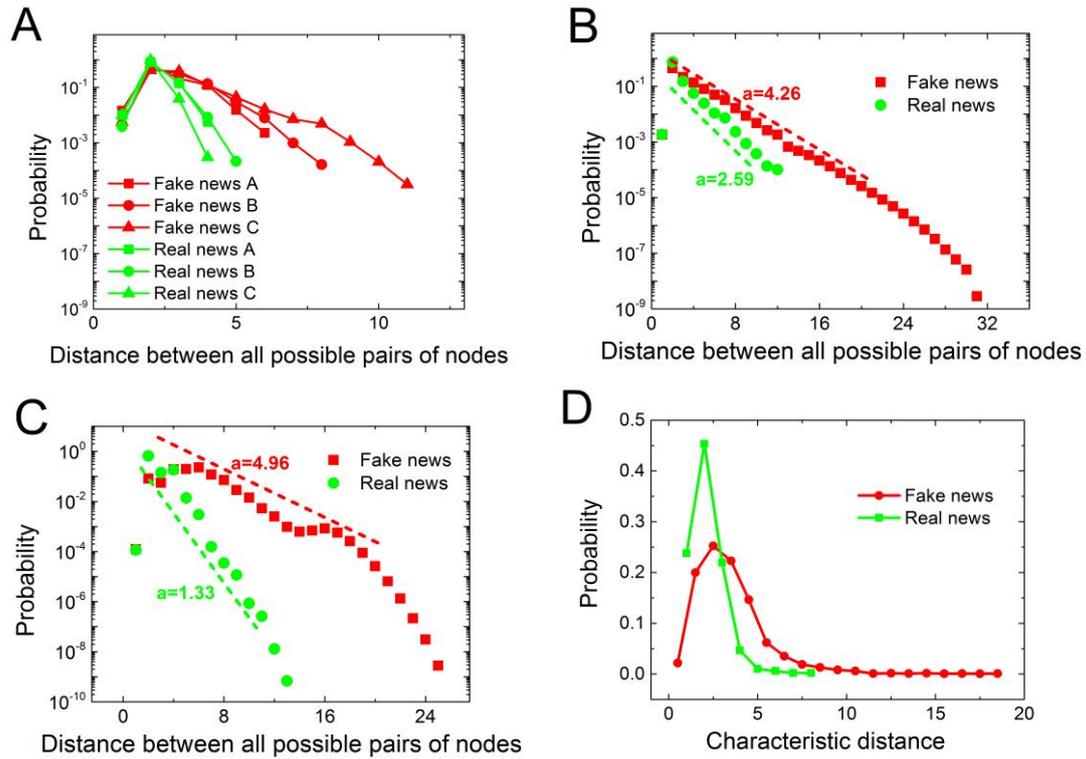

**Figure 4. Characteristic distance differentiates fake news from real news.** (A) The PDF of distances for three typical examples of networks for both fake and real news in Weibo. (B) The PDF of distances for all real and fake news networks in Weibo. (C) The PDF of distances of all real and fake news networks in Twitter. (D) The PDF of the characteristic distances (*details in methods*) for fake news and real news. The *p*-value between fake and real news is below 0.01.

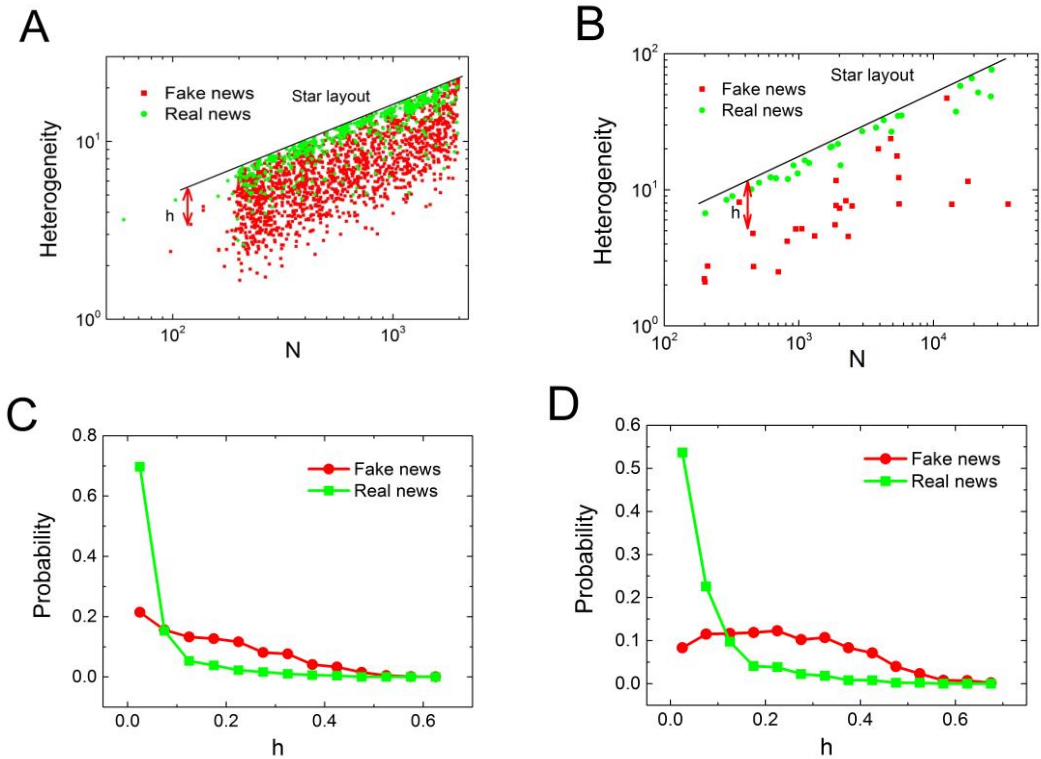

**Figure 5. Heterogeneity measure for fake and real news in Weibo and Twitter. (A)** The x axis is the size of the propagation network, and the y axis is the heterogeneity measure of the networks. The black line is the value of the star layout. The *h* is the difference of heterogeneity value between a real network and the corresponding value of star layout. (B) The scatter plot like in (A) for Twitter. (C) Distribution of *h* at five hours from the first re-posting of the Weibo propagation networks. The *p*-value here is below 0.01. (D) Distribution of *h* of all re-postings in Weibo for the whole lifespan. The *p*-value is below 0.01.

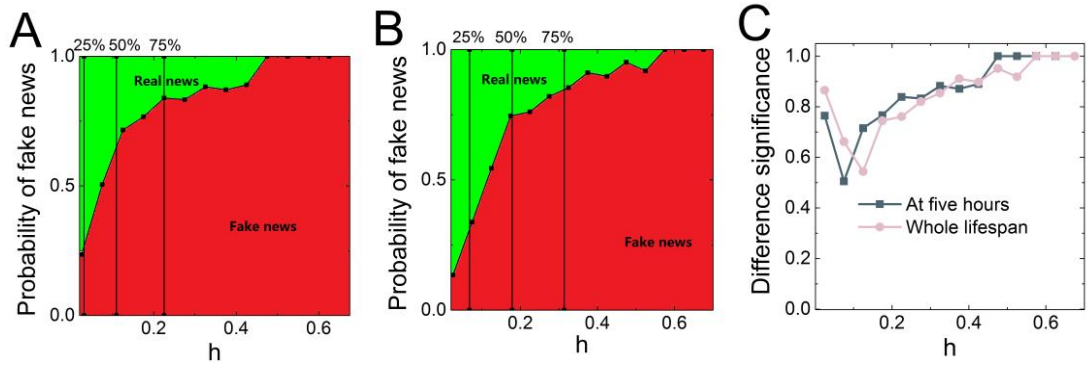

**Figure 6. Heterogeneity measure show high difference between fake news and real news of Weibo in its early stage of propagation.** (A) Probability of being fake news at five hours. The three vertical lines divide the figure into four parts with equal number of networks. For example, the area on the left of the first left line has 25% of all the Weibo propagation networks. (B) Probability of being fake news for the whole lifespan. (C) The difference significance between fake news and real news for the whole lifespan.

# Tables

**Table 1: Number of users and networks for different propagation networks**

|  | Fake news | Real news |
|---|---|---|
| Weibo users in the whole dataset | 973,391 | 347,401 |
| Twitter users in the whole dataset | 105,335 | 133,109 |
| Number of Weibo propagation networks (larger than 200 re-postings) | 1701 | 492 |
| Number of Weibo propagation networks which can be studied at five hours (larger than 200 re-postings) | 1578 | 448 |
| Number of Twitter propagation networks (larger than 200 re-postings) | 27 | 28 |

**Table 2: Comparison between three methods**

| Measure | Difference significance of Weibo | | |
|---|---|---|---|
|  | Whole lifespan | Five hours | Non-official creator |
| Ratio of layer sizes | 75.0% | 74.4% | 74.0% |
| Characteristic distance | 74.6% | 73.0% | 70.1% |
| Heterogeneity(h) | 78.2% | 74.1% | 71.3% |

*The Weibo networks are from the first Weibo datasets in these two tables.